\documentclass[traditabstract]{aa}
\usepackage{graphicx}
\usepackage{amsmath,amsfonts}

\def\Na{\boldsymbol{\nabla}}
\def\Nas{\boldsymbol{\nabla}_{s}}

\def\div{\boldsymbol{\nabla}\cdot}
\def\rot{\boldsymbol{\nabla}\times}
\newcommand{\abs}[1]{|#1|}
\newcommand{\Abs}[1]{\left|#1\right|}
\newcommand{\disc}[1]{[\![#1]\!]}

\def\n{{\bf{\hat n}}}
\def\tu{\hat{\bf t}}
\def\xu{{\bf{\hat x}}}
\def\yu{{\bf{\hat y}}}
\def\zu{{\bf{\hat z}}}
\def\ru{\boldsymbol{\hat{r}}}
\def\vphu{{\boldsymbol{\hat{\varphi}}}}
\def\uu{{\bf{\hat{u}}}}

\newcommand{\dep}[1]{\partial_{#1}}
\def\d{{\rm d}}
\def\dl{\,{\rm d} \mathit{l}}
\def\B{{\bf B}}
\def\Bz{B_{z}} 
\def\Bn{B_{n}}
\def\E{{\bf E}}
\def\En{E_{n}}
\def\v{{\bf v}}
\def\r{{\bf r}}
\def\bX{{\bf X}}
\def\j{{\bf j}}

\begin{document}

\title{A stability property of a force-free surface \\
 bounding a vacuum gap}
 
\titlerunning{}
 
\author{J.-J. Aly}

\offprints{J.-J. Aly}

 \institute{AIM - Unit\'e Mixte de Recherche CEA - CNRS - Universit\'e Paris VII 
              - UMR n$^{\tiny{0}}$ 7158, Centre d'Etudes de Saclay, F-91191  Gif sur Yvette Cedex, France;
              \email{jean-jacques.aly@cea.fr}}

\date{Published; A\&A 429, 779-784 (2005)}

\abstract{
A force-free surface (FFS) ${\cal S}$ is a sharp boundary separating
a void from a region occupied by a charge-separated 
force-free plasma. It is proven here under 
very general assumptions that there is 
on ${\cal S}$ a simple relation between the charge density $\mu$
on the plasma side and the derivative
of $\delta=\E\cdot\B$ along $\B$ on the vacuum side (with $\E$ denoting the electric field and $\B$ the magnetic field). 
Combined with the condition $\delta=0$ on ${\cal S}$, this relation implies
that a FFS has a general stability property, already conjectured by Michel (1979, ApJ 227, 579): ${\cal S}$ turns out to attract charges placed on the vacuum side if they 
are of the same sign as $\mu$. In the particular case of a FFS existing in the axisymmetric stationary magnetosphere of a ``pulsar'', the relation is given a most convenient form by using magnetic coordinates, and is shown to imply an interesting property of a gap.
Also, a simple proof is given of the impossibility of a vacuum gap forming in a field
$\B$ which is either uniform or radial (monopolar). 
}

\maketitle

\section{Introduction}

An important feature of many models of pulsar magnetosphere is the 
presence of voids or gaps embedded in a charge-separated force-free 
plasma (Holloway \cite{Hol1973}, Ruderman \& Sutherland \cite{RudSut1975},
Rylov \cite{Ryl1977}, Michel \cite{Mic1979}, Jackson
\cite{Jac1980}).
Such a void exists in particular in the global axisymmetric 
model first proposed by Michel (\cite{Mic1980}) and later numerically computed by several groups (Krause-Polstorff \& Michel 
\cite{KPM1985a,KPM1985b}, Smith et al. \cite{SMT2001}, P\'etri et al. \cite{PHB2002a},
Spitkovsky \& Arons \cite{SA2002}). In this latter picture, the void extends to infinity, the plasma occupying only
a bounded region -- the electrosphere -- around the pulsar.  
\par
  
In his general discussion of vacuum gaps, Michel (\cite{Mic1979}) reported some 
explicit examples of such structures, and noted that they have all a 
remarkable property. On the surface of discontinuity ${\cal S}$ separating the 
plasma of positive (negative) charges from the vacuum -- a so-called 
force-free surface (FFS) --, the component of the electric field
along the magnetic field on the vacuum side is always directed in such a way that positive (negative) particles be attracted by ${\cal S}$. 
Michel made the guess that this phenomenon is general, the 
existence of a FFS thus implying some form of stability. It is
the aim of this paper to give a complete analytical proof of this
conjecture. Actually, we shall establish its validity under
assumptions much weaker than Michel's, as we shall assume 
neither that the system is stationary and axisymmetric, nor that 
the magnetic field is unaffected by the local electric currents.
\par
 
The paper is organized as follows. In Sect. 2, we consider a 
general continuous electromagnetic field $(\E,\B)$ having its gradient 
$\Na(\E,\B)$ suffering a discontinuity accross 
a moving surface ${\cal S}$, and establish 
a jump relation for the normal derivative of the scalar product $\E\cdot\B$.
Our starting point here is the complete system of Maxwell equations, and a set
of jump relations for $\Na(\E,\B)$ which are established in Appendix A. In Sect. 3, the jump relation for $\dep{n}(\E\cdot\B)$ is applied to the case where ${\cal S}$ is a FFS, and it is shown to imply a general formula from which the validity of Michel's conjecture follows immediately. Sect. 4 is devoted to the case of a FFS existing in the axisymmetric stationary magnetosphere of a pulsar. We show in particular how the formula derived in Sect. 3 can be conveniently rewritten by using magnetic coordinates and derive a general property of gaps. Incidentally, we give in Appendix B (in addition to a simple explicit example illustrating our general relation) a proof of the impossibility of a void bounded by a FFS and threaded by a field $\B$ which is either uniform or radial.

\section{A general discontinuity relation}

Let us consider a domain $D$ of space in which there is a moving 
smooth surface ${\cal S}={\cal S}(t)$. We choose a unit normal $\n$ on ${\cal S}$, and define its ``$+$'' side as the one towards which $\n$ is 
pointing and its ``$-$'' side as the other one. For any function 
$X$ (scalar or vector) continuous in $D\setminus{\cal S}$ and having well 
defined values on ${\cal S}^{\pm}$, we define the jump 
$\disc{X}$ on ${\cal S}$ to be
\begin{eqnarray}
 	\disc{X} = X^{+} - X^{-} \,.
 	\label{DefDisc}
\end{eqnarray}
On ${\cal S}$, we introduce the decompositions 
\begin{eqnarray}
	{\bX} = {\bX}_{s}+X_{n}\n 
	\label{DecXS}
\end{eqnarray}
for any vector field $\bX$, and
\begin{eqnarray}
	\Na = \Nas + \n\,\dep{n}  
	\label{DecNaS}
\end{eqnarray}
for the nabla operator, with $\Nas$ acting parallel to ${\cal S}$ and $\dep{n}=\n\cdot\Na$ being the 
derivative along $\n$. Moreover, in order to define 
the normal derivatives of ${\bX}_{s}$ and $X_{n}$ (which are yet defined only on ${\cal S}$), we extend the vector $\n$ -- and accordingly the decomposition (\ref{DecXS}) -- 
into a thin layer around ${\cal S}$ by imposing $\dep{n}\n=\n\cdot\Na\n=0$ (this amounts to set $\n(\r)=\n(\r_{p})$, with $\r_{p}$ the projection of $\r$ onto ${\cal S}$). Note that $\Nas$ and $\dep{n}$ do not commute in general. But obviously $\Nas$ 
commutes with $\disc{.}$, i.e.,
\begin{eqnarray}
	\disc{\Nas X} = \Nas\disc{X} \,.
	\label{NasDisc}
\end{eqnarray}
\par

Let us now assume that $D$ contains an electromagnetic field $(\E,\B)$ such that:
\par

\begin{itemize}
        \item 
        $(\E,\B)$ is continuous in $D$, whence in particular
        \begin{eqnarray}
             \disc{\E} = \disc{\B} = 0 
             \;\;\;\;\; \mbox{on } \;\; {\cal S} \,.
             \label{ContEBS}
       \end{eqnarray}
             
       \item
       $(\E,\B)$ is continuously differentiable in $D\setminus{\cal S}$, with
       $\Na(\E,\B)$ being well defined on ${\cal S}^{-}$ and 
       ${\cal S}^{+}$, and $\disc{\Na(\E,\B)}$ thus being well defined
       on ${\cal S}$.
       
       \item
       In $D\setminus{\cal S}$, $(\E,\B)$ does satisfy Maxwell equations
       \begin{eqnarray}
  	      c\rot\E &=& -\dep{t}\B \,,
  	      \label{Max1} \\
  	      \div\B &=& 0 \,,
  	      \label{Max2}  \\
  	      \div\E &=& 4\pi\mu \,,
  	      \label{Max3} \\
  	      c\rot\B &=& 4\pi\j + \dep{t}\E \,,
  	      \label{Max4}
       \end{eqnarray}
       where $\mu$ and $\j$ are the electric charge and 
       current densities, respectively, and $c$ is the speed of light.
\end{itemize}
\par

Under these assumptions, it is  possible to derive the following set of jump relations for $\Na(\E,\B)$:      
\begin{eqnarray}
	\disc{\Nas\E} &=& 0 \,, 
 	\label{DiscNasE}  \\ 	
       \disc{\Nas\B} &=& 0 \,, 
 	\label{DiscNasB}  \\
	\disc{\dep{n}\En} &=& 4\pi\disc{\mu} \,,
	\label{DiscdnEn}   \\
	\disc{\dep{n}\Bn} &=& 0 \,,
	\label{DiscdnBn}   \\
 	\disc{\partial_{n}\E_{s}} &=& -\frac{4\pi}{c^{2}}\gamma_{n}^{2}V_{n}
 	\disc{\j_{s}} \,,
 	\label{DiscdnEs}  \\
	\disc{\partial_{n}\B_{s}} &=& \frac{4\pi}{c}\gamma_{n}^{2}
 	\disc{\j_{s}}\times\n \,,
 	\label{DiscdnBs} 
\end{eqnarray}
where 
 \begin{eqnarray}
 	\gamma_{n}^{2} = \left(1-\frac{V_{n}^{2}}{c^{2}}\right)^{-1}
 	\label{Defgamn}
 \end{eqnarray}
is the squared Lorentz factor associated with the normal velocity $V_{n}$ of ${\cal S}$. Moreover, the discontinuity of the normal component $j_{n}$ of the current and that of the charge density $\mu$ have to satisfy the constraint
\begin{eqnarray}
	\disc{j_{n}} &=& \disc{\mu}V_{n} \,.
	\label{RelDiscjmu} 
\end{eqnarray}
Quite certainly, Eqs. (\ref{DiscNasE})-(\ref{DiscdnBs}) and (\ref{RelDiscjmu}) have
already been established and used in many other contexts, but we 
are not aware of any references, and then we give a proof of them in Appendix A.
\par

We are now ready to write the general relation 
expressing the discontinuity of 
the normal derivative
\begin{eqnarray}
	\disc{\dep{n}(\E\cdot\B)}
	&=& \E_{s}\cdot\disc{\dep{n}\B_{s}}
	+ E_{n}\disc{\dep{n}B_{n}}
       \nonumber \\
	&& + \B_{s}\cdot\disc{\dep{n}\E_{s}}
	+ B_{n}\disc{\dep{n}E_{n}}
	\label{¥}
\end{eqnarray}
of the scalar product $\E\cdot\B$
as a function of $V_{n}$ and of the discontinuities of 
$\j$ and $\mu$. Using equations (\ref{DiscdnEn})-(\ref{DiscdnBs}), we obtain
\begin{eqnarray}
	\disc{\dep{n}(\E\cdot\B)}
	&=& 4\pi\disc{\mu}B_{n}
       \nonumber \\
	&& + \frac{4\pi}{c}\gamma_{n}^{2}
	\left(\n\times\E_{s}-\frac{V_{n}}{c}\B_{s}\right)\cdot\disc{\j_{s}} \,.
	\label{DiscdnE.B} 
\end{eqnarray}

\section{Application to a force-free surface}

\subsection{Assumptions}

We now assume that ${\cal S}$ is a FFS, i.e., 
there is a charge-separated force-free plasma on 
one side of ${\cal S}$ -- ${\cal S}^{-}$, say -- and a vacuum on 
the other side. 
We thus have the additionnal bulk relations
\begin{eqnarray}
	\j &=& \mu\v \,,
	\label{jChSepReg} \\
	\E &=& -\frac{\v\times\B}{c} \,,
	\label{Ohm}
\end{eqnarray}
inside the plasma,
and 
\begin{eqnarray}
       \j &=& 0 \,,
       \label{jVacuum} \\
       \mu &=& 0 \,,
       \label{muVacuum}
\end{eqnarray}
inside the vacuum. This implies
\begin{eqnarray}
	\disc{\mu} &=& -\mu^{-} \,,
	\label{DiscmuFFS} \\
	\disc{\j} &=& -\mu^{-}\v^{-} \,,
	\label{DiscjsFFS}Ê\\
	\E_{s} &=& -B_{n}\frac{\v_{s}^{-}}{c}\times\n
	- \frac{V_{n}}{c}\n\times\B_{s} 
	\label{EsPlFFS}
\end{eqnarray}
on ${\cal S}$, where we have noted to get the last relation that,
as a consequence of Eqs. (\ref{RelDiscjmu}) and (\ref{DiscmuFFS})-(\ref{DiscjsFFS}),
$v_{n}^{-}= V_{n}$ --
this equality just expressing the fact that ${\cal S}$ stays a 
plasma-vacuum interface at any time.

\subsection{The discontinuity relation}

Hereafter, we denote as $\ell$ an arclength along a magnetic line, with
$\ell$ increasing
when passing from the plasma to the vacuum, and as $\uu$ the associated tangent
unit vector (then $\uu\cdot\n\geq 0$ on ${\cal S}$). Our aim is to compute the value of the derivative with respect to $\ell$ of the component
\begin{eqnarray}
	E_{\parallel} = \E\cdot\uu 
	\label{¥}
\end{eqnarray}
of the electric field.
\par

As $\E\cdot\B=0$ on the plasma side, we have
on ${\cal S}$
\begin{eqnarray}
	\E\cdot\B &=& 0 \,, 
	\label{EBFFS} \\
	\left[\Nas(\E\cdot\B)\right]^{+} &=& 0 \,,
	\label{FFS-dsEB} \\
	\disc{\dep{n}(\E\cdot\B)} &=& [\dep{n}(\E\cdot\B)]^{+} \,.
	\label{FFS-dnEB}
\end{eqnarray}
Then $E_{\parallel}$ satisfies on ${\cal S}$
\begin{eqnarray}
	E_{\parallel}^{+} = E_{\parallel} =  0 \,,
	\label{EparVac}
\end{eqnarray}
and
\begin{eqnarray}
    \left.B^{2}\frac{\d E_{\parallel}}{\d\ell}\right|^{+}
    &=& [\B\cdot\Na(\E\cdot\B)]^{+}
    \nonumber \\
    &=& [\B_{s}\cdot\Nas(\E\cdot\B)]^{+} 
    + [\Bn\dep{n}(\E\cdot\B)]^{+} 
    \nonumber \\
    &=& \Bn\disc{\dep{n}(\E\cdot\B)} \,,
	\label{dlEpar}
\end{eqnarray}
where we have used equations (\ref{FFS-dsEB})-(\ref{FFS-dnEB}) 
to get the last equality.
Using Eqs. (\ref{DiscmuFFS})-(\ref{EsPlFFS})  in 
the right-hand side of equation (\ref{DiscdnE.B}) and remembering that 
$v_{n}^{-}=V_{n}$, we get 
\begin{eqnarray}
	\disc{\dep{n}(\E\cdot\B)}
	&=& -4\pi\mu^{-}B_{n}\left(1-\gamma_{n}^{2}
	\frac{(v_{s}^{-})^{2}}{c^{2}}\right)
	\nonumber \\
       &=& -4\pi\mu^{-}B_{n}\gamma_{n}^{2}\gamma^{-2} \,,
	\label{¥}
\end{eqnarray}
where 
\begin{eqnarray}
	\gamma^{-2} = 1-\frac{(v_{n}^{-})^{2}+(v_{s}^{-})^{2}}{c^{2}} \,.
	\label{¥}
\end{eqnarray}
\par

By using the latter relation in equation (\ref{dlEpar}), we obtain finally
\begin{eqnarray}
	\left.\frac{\d E_{\parallel}}{\d\ell}\right|^{+}
	= -4\pi\mu^{-}\frac{B_{n}^{2}}{B^{2}}
	\gamma_{n}^{2}\gamma^{-2} \,,
	\label{¥}
\end{eqnarray}
or, by introducing the angle $\alpha$ between $\B$ and $\n$, 
\begin{eqnarray}
	\left.\frac{\d E_{\parallel}}{\d\ell}\right|^{+}
	= -4\pi\mu^{-}\cos^{2}\alpha\gamma_{n}^{2}\gamma^{-2} \,. 
	\label{dEpar/dl}
\end{eqnarray}
\par
       
On a part of ${\cal S}$ not tangent to $\B$ ($\cos\alpha\neq 0$), we thus have
\begin{eqnarray}
	\mu^{-}\left.\frac{\d E_{\parallel}}{\d\ell}\right|^{+}
	= -4\pi(\mu^{-}\cos\alpha\gamma_{n}\gamma^{-1})^{2} < 0 \,, 
	\label{}
\end{eqnarray}
which implies along with Eq. (\ref{EparVac}) the existence of a layer on the vacuum side in which
\begin{eqnarray}
       \mu^{-}E_{\parallel} < 0 \,,
       \label{}
\end{eqnarray}
i.e., the sign of $E_{\parallel}$ is opposite to that of $\mu^{-}$. 
When $\mu^{-}>0$, $E_{\parallel}$ is directed towards ${\cal S}$, which attracts the positive particles, while when $\mu^{-}<0$, $E_{\parallel}$ is directed 
away from ${\cal S}$, which attracts the negative ones. This is just the general stability property conjectured by Michel (\cite{Mic1979}).
\par

As an illustration of our general formula (\ref{dEpar/dl}), we consider in Appendix B a particular example of FFS  borrowed  from Michel \cite{Mic1989}.

\section{Application to the aligned pulsar magnetosphere}

\subsection{Assumptions and the basic discontinuity relation}

In this section, we use standard spherical coordinates $(r,\theta,\varphi)$ 
attached to the Cartesian frame $(O,\xu,\yu,\zu)$. We consider a pulsar of center $O$ rotating at the angular velocity $\Omega_{*}\zu$ and surrounded by a stationnary
magnetosphere $D$. The latter is axisymmetric around the $z$-axis -- this allows us to consider the equations governing its structure as being set in 
the meridional half-plane $\Pi=\{\varphi=0\}$ --, and it is made of two regions
separated by a FFS ${\cal S}$: 
$D_{e}$, which contains a charge-separated force-free plasma, and $D_{v}$ which is a vacuum. Here we allow
$D_{e}$ and $D_{v}$ to be constituted of several connected parts and/or to
extend to infinity
(we do not refer to any specific model). We denote as $\Gamma$ and $\partial\Gamma$ the intersections of  
$D_{v}$ and ${\cal S}$, respectively, with $\Pi$.   
\par

In the whole $D$: (i) the magnetic field admits the well known representation
\begin{eqnarray}
       \B = \frac{\Na A\times{\boldsymbol{\hat{\varphi}}}}{r\sin\theta} \,,
       \label{}
\end{eqnarray}
with the \textit{flux function} $A$ satisfying
\begin{eqnarray}
	\Na^{2}A - 2\Bz = -\frac{4\pi}{c}\mu(r\sin\theta)^{2}\Omega
	\label{EqA}
\end{eqnarray}
in $D_{v}$ (with $\mu=0$) and in the part of $D_{e}$  where the particles do not stream along the lines and then have a purely rotational motion at the angular velocity $\Omega$; (ii) the electric field can be written in the form
\begin{eqnarray}
	\E = - \Na\psi 
	\label{}
\end{eqnarray}
owing to the stationarity assumption,
with the electrostatic potential $\psi$ solving  Poisson equation
\begin{eqnarray}
	-\Na^{2}\psi = 4\pi\mu \,.
	\label{Eqpsi}
\end{eqnarray}
Inside the plasma, $\psi=\psi(A)$, with 
\begin{eqnarray}
	\psi_{A}^{\prime}(A) = \frac{\d\psi}{\d A} = \frac{\Omega}{c} 
	\label{ExprOmega}
\end{eqnarray}
(see, e.g., Goldreich \& Julian \cite{GJ1969}).
\par 

Let us write the particular form taken by Eq. (\ref{dEpar/dl}) on $\partial\Gamma$. As ${\cal S}$ does not move:  (i) $V_{n}=0$ and $\gamma_{n}=1$; (ii) 
a charged element on ${\cal S}$ has a pure rotational motion at the angular velocity 
$\Omega$ given by Eq. (\ref{ExprOmega}) ($\Omega=\Omega_{*}$ if the element is connected to the star by a piece of magnetic line fully embedded in $D_{e}$).
We thus have
\begin{eqnarray}
	\left.\frac{\d E_{\parallel}}{\d\ell}\right|^{+}
	= -\left.\frac{\d^{2}\psi}{\d\ell^{2}}\right|^{+}
	= -4\pi\mu^{-}\gamma^{-2}\cos^{2}\alpha \,, 
	\label{dEpar/dlAx}
\end{eqnarray}
with
\begin{eqnarray}
	\gamma^{-2} = 1-(r\sin\theta\Omega/c)^{2} \,. 
	\label{gammaAx}
\end{eqnarray}

\subsection{Expression of the discontinuity relations in magnetic coordinates}

Inside the vacuum region $D_{v}$, the magnetic field also admits
the representation
\begin{eqnarray}
       \B = \Na V \,,
       \label{}
\end{eqnarray}
with the scalar potential $V$ satisfying 
\begin{eqnarray}
	\Na^{2}V &=& 0 \,.
\end{eqnarray}
It is often convenient to use $A$ and $V$
as curvilinear orthogonal coordinates in $\Gamma$ 
(note that $\Na A\cdot\Na V=0$). The vacuum electrostatic potential written as a function $\psi(A,V)$ thus satisfies 
\begin{eqnarray}
       \Na\psi &=& \psi^{\prime}_{A}\Na A + \psi^{\prime}_{V}\Na V \,,
       \label{NapsiAx} \\
       \E\cdot\B &=& -\psi^{\prime}_{V}B^{2} \,,
       \label{EBAx} \\
       \Na^{2}\psi &=& \psi^{\prime\prime}_{AA}\abs{\Na A}^{2}
       + \psi^{\prime\prime}_{VV}\abs{\Na V}^{2} + \psi_{A}^{\prime}\Na^{2}A
	\nonumber \\
	&=& [(r\sin\theta)^{2}\psi^{\prime\prime}_{AA} 
       + \psi^{\prime\prime}_{VV}]B^{2} + 2\psi_{A}^{\prime}\Bz
       = 0 
       \label{Na2psiAx}
\end{eqnarray}
(Eq. (\ref{EqA}) has been used to get the last line). 
\par

By Eqs. (\ref{EBAx}) and (\ref{EBFFS}), we have on $\partial\Gamma$
\begin{eqnarray}
       \psi^{\prime}_{V} = 0 \,,
       \label{dpsiV}
\end{eqnarray}
while Eqs. (\ref{EBAx}) and (\ref{dEpar/dlAx}) imply
\begin{eqnarray}
	4\pi\mu^{-}\gamma^{-2}\cos^{2}\alpha B^{2}
	= -\B\cdot[\Na(\E\cdot\B)]^{+} 
	\nonumber \\
	= \B\cdot[\Na(B^{2}\psi^{\prime}_{V})]^{+}
	= B^{4}\psi^{\prime\prime}_{VV}\mid^{+} ,
	\label{}
\end{eqnarray}
whence
\begin{eqnarray}
	\psi^{\prime\prime}_{VV}\mid^{+} 
	= \frac{4\pi\mu^{-}\gamma^{-2}\cos^{2}\alpha}{B^{2}} \,.
	\label{dpsiVV}
\end{eqnarray}
It is interesting to note that the two other second derivatives of $\psi(A,V)$ can be also given explicit expressions. To get $\psi^{\prime\prime}_{AV}$, we differentiate Eq. 
(\ref{dpsiV}) along 
$\partial\Gamma$. Denoting by $l$ an arclength increasing in the direction of the  tangent unit vector  
$\tu=\boldsymbol{\hat{\varphi}}\times\n$
and with $\alpha$ being oriented by $\vphu$, we have
\begin{eqnarray}
	0 &=& \psi^{\prime\prime}_{AV}\mid^{\pm}\frac{\d A}{\dl}
	+ \psi^{\prime\prime}_{VV}\mid^{\pm}\frac{\d V}{\dl}
	\nonumber \\
	&=& \psi^{\prime\prime}_{AV}\mid^{\pm}r\sin\theta\cos\alpha B
	+ \psi^{\prime\prime}_{VV}\mid^{\pm}\sin\alpha B \,,
	\label{}
\end{eqnarray}
and then
\begin{eqnarray}
	\psi^{\prime\prime}_{AV}\mid^{+} 
	= -\frac{4\pi\mu^{-}\gamma^{-2}\sin\alpha\cos\alpha}{r\sin\theta B^{2}} \,.
	\label{dpsiAV}
\end{eqnarray}
And by using Eq. (\ref{Na2psiAx}), we obtain
\begin{eqnarray}
	\psi^{\prime\prime}_{AA}\mid^{+} 
	&=&  \frac{-1}{(r\sin\theta B)^{2}}\left(
	4\pi\mu^{-}\gamma^{-2}\cos^{2}\alpha+2\psi_{A}^{\prime}\Bz\right)
	\nonumber \\
	&=& \frac{-4\pi}{(r\sin\theta B)^{2}}\left(
	\mu^{-}\gamma^{-2}\cos^{2}\alpha
	- \mu_{*}\gamma_{*}^{-2}\frac{\Omega}{\Omega_{*}}\right) ,
	\label{dpsiAA}
\end{eqnarray}
where $\mu_{*}=-\Omega_{*}\Bz\gamma_{*}^{2}/2\pi c$
is the Goldreich-Julian or corotational charge density (Goldreich \& Julian
\cite{GJ1969}) and $\gamma_{*}$ is given by Eq. (\ref{gammaAx}) with $\Omega=\Omega_{*}$.  
\par

Near the pulsar, there is a region $D'$ where the particles move at a velocity small compared to $c$ and the magnetic field is essentially determined by 
the sole currents flowing inside the star (therefore both $A$ and $V$ are a priori known quantities in $D'$, and using them as coordinates therein seems to be especially appropriate). In many papers, the authors consider a FFS which is located in such a region (this is the case in all the calculations of Michel's electrosphere referred to in the introduction) and it is  possible to take $\gamma=1$ in all the relations above. In that situation,
a simplified proof of Eqs. (\ref{dpsiVV}), (\ref{dpsiAV}) and (\ref{dpsiAA}) can be obtained by making the nonrelativistic approximation from the very beginning. In fact, it is just needed to use Eqs. (\ref{EqA}) and 
(\ref{Eqpsi}) (the former with $\mu=0$ everywhere) and to take into account the continuity of $\psi$ and $\Na\psi$ on ${\cal S}$.

\subsection{A general property of gaps}

Let us define a {\it vacuum force-free surface} (VFFS) to be a surface located inside 
a void on which $\E\cdot\B=0$. Then we have the following result: there is necessarily such a surface in $D_{v}$ if the latter contains a piece of a magnetic line connecting two points $\r_{1}$ and $\r_{2}$ of $\partial\Gamma$ at which the charge density has the same sign -- i.e., $\mu(\r_{1})\mu(\r_{2})>0$ -- and $\cos\alpha\neq 0$.
\par
 
The proof of this statement is immediate. As a consequence of Eqs. (\ref{EBFFS}) 
and (\ref{dEpar/dl}), we have
\begin{eqnarray}
	\delta(\r_{1}) = \delta(\r_{2}) = 0 \,,
	\label{}
\end{eqnarray}
where $\delta=\E\cdot\B$, and 
\begin{eqnarray}
	(\B\cdot\Na\delta)(\r_{1})\cdot(\B\cdot\Na\delta)(\r_{2})
	< 0 \,,
	\label{}
\end{eqnarray}
which clearly implies that $\delta$ has to vanish at least once (in fact $2p+1$ times, with $p\geq 0$ an integer) on the line. 
By continuity, the same result holds on all the neighboring lines, and the points where $\delta=0$ form a surface indeed -- a VFFS. 
\par

A trivial illustration of this result is provided by the calculations of Michel's electrospheric structure in a purely dipolar field. In that case, there is a bundle
of magnetic lines connecting through  the unbounded vacuum region the negatively charged domes overlying the north and south poles, respectively,
and it is a priori quite obvious that this bundle threads a VFFS 
which is just the part of the equatorial plane it does intersect -- we have indeed $\E\cdot\B=0$ on $\{z=0\}\cap D$ as a consequence of the symmetry of the system.
\par

As a more interesting illustration, we may consider -- following Ass\'{e}o et al. (\cite{ABP1984}) -- the theoretical possibility of a connected gap $d_{v}$ fully included in the negatively charged region overlying the North pole, say. If such a gap does exist, it necessarily contains by the result above a 
VFFS ${\cal S}_{v}$ (possibly made of several parts) cutting all the magnetic lines an odd number of times. In particular, ${\cal S}_{v}$ meets the $z$-axis at a critical point of $\psi$ ($\Na\psi=0$) where at least two equipotential surfaces meet (this point has to be of the $X$-type as the vacuum potential cannot reach either a maximum or a minimum inside $d_{v}$). As $\E$ also vanishes at the points of intersection of the boundary of $d_{v}$ with the $z$-axis, with each of these points generating a cusped equipotential surface (Ass\'{e}o et al. \cite{ABP1984}), we see that the potential needs to have a quite curious structure, which makes the existence of $d_{v}$ unlikely (this point is discussed in more details in a forthcoming paper, where we combine the result here with other properties of the potential $\psi$ to recover Ass\'{e}o et al.'s nonexistence result without appealing to physical arguments involving pair production). 
\par

Incidentally we note that the impossibility of a stationary vacuum gap (either of finite extent, or unbounded with $\E$ vanishing at infinity) can be most easily proved in the nonrelativistic approximation when the given potential field $\B$ is uniform or radial (see Appendix B). An hypothetical pulsar with a monopolar (or ``split monopolar'') magnetic field could then not have a finite electrosphere of Michel's type. It is worth noticing that Michel (\cite{Mic1973}) has found  in that situation an exact gapless solution of the Goldreich-Julian type, with a wind extending to infinity.

\section{Conclusion}

In this paper, we have provided a general analytical proof of a stability conjecture put forward by Michel (\cite{Mic1979}) some twenty five years ago. According to the latter, any FFS does attract particles located on the vacuum side if they are of the same sign as the particles on the plasma side, a property which makes the electrostatic at work in a pulsar magnetosphere with gaps somewhat nonstandard. Although Michel formulated his conjecture on the basis of particular axisymmetric, stationary and nonrelativistic examples, we have shown here that it was quite robust, still holding true when neither of these simplifying assumptions are made. 
\par

In the case where the FFS exists in the magnetosphere of an ``aligned pulsar'',
we have shown that the general relation from which we have derived the validity of Michel's conjecture takes a particularly simple form when magnetic coordinates are used. This relation determines indeed the second derivative of the vacuum electrostatic potential $\psi$ with respect to the magnetic potential $V$ as a function of the charge density, 
the condition that ${\cal S}$ be a FFS being expressed by the vanishing of the first derivative of $\psi$ with respect to that same variable.
We have also derived an interesting property of a vacuum gap threaded by magnetic lines connecting plasma elements having the same sign of $\mu$ -- such a gap does necessarily contain a VFFS -- and proved the impossibility to have in a given uniform or radial magnetic field a bounded stationary gap, or an unbounded one with no electric sources at infinity.

\appendix

\section{Jump relations for $\Na(\E,\B)$}

We give here a derivation of the jump relations for $\Na(\E,\B)$ quoted in Section 2.
\par

\begin{enumerate}
	\item
	\textit{Derivation of Eqs. (\ref{DiscNasE})-(\ref{DiscNasB})}. 
	As an immediate consequence of 
	Eqs. (\ref{NasDisc})-(\ref{ContEBS}), we have
	\begin{eqnarray}
 		(\disc{\Nas\E},\disc{\Nas\B}) = (\Nas\disc{\E},\Nas\disc{\B}) = (0,0) \,. 
 		\label{DiscNasEBAp}
	\end{eqnarray}
	
	\item
	\textit{Derivation of Eqs. (\ref{DiscdnEn})-(\ref{DiscdnBn})}.
	By the standard decomposition
	\begin{eqnarray}
		\div\bX = \Nas\cdot\bX + \dep{n}X_{n}
		\label{}
	\end{eqnarray}
	of the divergence operator on a 
	surface (e.g., Brand \cite{Bra1947}) and Eq. (\ref{DiscNasEBAp}),
	we have 
	\begin{eqnarray}
  		\disc{\div(\E,\B)} 
  		&=& \disc{\Nas\cdot(\E,\B)}
  		+ \disc{\dep{n}(E_{n},B_{n})} 
  		\nonumber  \\
       		&=& (\disc{\dep{n}E_{n}},\disc{\dep{n}B_{n}}) \,.
  		\label{Discdiv}
	\end{eqnarray}
	On the other hand, we have by Eqs. (\ref{Max2}) and (\ref{Max3})
	\begin{eqnarray}
  		\disc{\div\B} &=& 0 \,,
  		\label{DiscMax2a}  \\
  		\disc{\div\E} &=& 4\pi\disc{\mu} \,,
  		\label{DiscMax3a}
	\end{eqnarray}
	and then 
	\begin{eqnarray}
		\disc{\dep{n}\Bn} &=& 0 \,,
		\label{DiscdnBnAp}   \\
		\disc{\dep{n}\En} &=& 4\pi\disc{\mu} \,.
		\label{DiscdnEnAp} 
	\end{eqnarray}

	\item
	\textit{Derivation of Eqs. (\ref{DiscdnEs})-(\ref{DiscdnBs}) and (\ref{RelDiscjmu})}.
	We note that Eq. (\ref{ContEBS}) implies
	\begin{eqnarray}
		\d_{t}\disc{\E,\B} = \disc{\dep{t}(\E,\B)} + 
		V_{n}\disc{\dep{n}(\E,\B)} = 0 \,,
		\label{DiscCinem}
	\end{eqnarray}
	where $\d_{t}$ is the time derivative 
	following the normal motion of ${\cal S}$, and we use
	the decomposition 
	\begin{eqnarray}
		\rot\bX = \Nas\times\bX + \n\times\dep{n}\bX_{s}
		\label{}
	\end{eqnarray}
	of the rotational operator on a surface (Brand \cite{Bra1947})
	and Eq. (\ref{DiscNasEBAp}) to write
	\begin{eqnarray}
		\disc{\rot(\E,\B)} &=& \disc{\Nas\times(\E,\B)} 
		+ \n\times\disc{\dep{n}(\E_{s},\B_{s})}
		\nonumber \\
      		&=& (\n\times\disc{\dep{n}\E_{s}},\n\times\disc{\dep{n}\B_{s}}) \,.
		\label{DiscRot}
	\end{eqnarray}
	Plugging these relations into the ones
	\begin{eqnarray}
  		c\disc{\rot\E} &=& -\disc{\dep{t}\B} \,,
  		\label{DiscMax1a} \\
  		c\disc{\rot\B} &=& 4\pi\disc{\j} + \disc{\dep{t}\E} \,,
  		\label{DiscMax4a}
	\end{eqnarray}
	 resulting from Eqs. (\ref{Max1}) and (\ref{Max4}) thus leads to
	\begin{eqnarray}
  		c \n\times\disc{\dep{n}\E_{s}} &=& V_{n}\disc{\dep{n}\B} \,,
  		\label{DiscMax1b} \\
  		c \n\times\disc{\dep{n}\B_{s}} &=& 4\pi\disc{\j} 
  		- V_{n}\disc{\dep{n}\E} \,.
  		\label{DiscMax4b}
	\end{eqnarray}
	The normal component of equation (\ref{DiscMax1b}) gives once more 
	Eq. (\ref{DiscdnBnAp}), 
	while its tangential component is equivalent to
	\begin{eqnarray}
  		\disc{\dep{n}\E_{s}} 
  		= \frac{V_{n}}{c}\disc{\dep{n}\B_{s}}\times\n \,.
  		\label{DiscMax1c} 
	\end{eqnarray}
	By taking similarly the normal and tangential component of
	Eqs. (\ref {DiscMax4b}), we obtain
	\begin{eqnarray}
		\disc{j_{n}} = \disc{\mu}V_{n} \,,
		\label{RelDiscjmuAp} 
	\end{eqnarray}
	where we have taken Eq. (\ref{DiscdnEnAp}) into account, and
 	\begin{eqnarray}
   		\n\times\disc{\partial_{n}\B_{s}} 
		= \frac{4\pi}{c}\disc{\j_{s}}
    		-\frac{V_{n}}{c}\disc{\dep{n}\E_{s}} \,.
		\label{DiscMax4d}
	\end{eqnarray}
	Finally, Eqs. (\ref{DiscMax1c}) and 
	(\ref{DiscMax4d}) combine to give
	\begin{eqnarray}
 		\disc{\partial_{n}\B_{s}} &=& \frac{4\pi}{c}\gamma_{n}^{2}
 		\disc{\j_{s}}\times\n \,,
 		\label{DiscdnBsAp}  \\
 		\disc{\partial_{n}\E_{s}} &=& -\frac{4\pi}{c^{2}}\gamma_{n}^{2}V_{n}
 		\disc{\j_{s}} \,,
 		\label{DiscdnEsAp}
	\end{eqnarray}
	with $\gamma_{n}$ defined by Eq. (\ref{Defgamn}). 
\end{enumerate}

\section{Particular configurations}

In this appendix, we present a brief study of three particular configurations. We assume stationarity -- then $\E$ can be expressed in terms of an electrostatic potential $\psi$ -- and nonrelativistic velocities, and take $\B$ to be a given potential field.  We use Cartesian coordinates $(x,y,z)$ and associated spherical coordinates $(r,\theta,\varphi)$.

\subsection{Spherical plasma confined by a uniform $\B$}

Following Michel \cite{Mic1989}, we consider a ball $V$ of center $O$ and radius $r_{0}$ containing free electric charges distributed with the uniform density $\mu$. $V$ is threaded by a uniform magnetic field $B\zu$ ($B>0$) and it is submitted to an external quadrupolar electric field 
oriented in such a way that the total electric field be given by
\begin{eqnarray}
	\E = \left\{ \begin{array}{lll}
	\frac{4\pi\mu r_{0}^{3}}{3r^{3}}\r + k(x\xu+y\yu-2z\zu) & \mbox{for } 
	& r\geq r_{0} \,,
	\\
	\\
	\frac{4\pi\mu}{3}\r + k(x\xu+y\yu-2z\zu) 
	& \mbox{for } & r\leq r_{0} \,.
	\label{}
	\end{array} \right.
\end{eqnarray}
From now on, we choose the parameter $k=2\pi\mu/3$. Then
$\E\cdot\B=BE_{z}=0$ in $V$, and the plasma turns out to be force-free if we take it to move at the drift velocity
\begin{eqnarray}
	\v = c\frac{\E\times\B}{B^{2}} = -\frac{2\pi cr\sin\theta\mu}{B}\vphu 
	\label{}
\end{eqnarray}
corresponding to a rigid rotation at the angular velocity
$\Omega=-2\pi c\mu/B$. Of course, $E_{z}=0$ on the sphere $\{r=r_{0}\}$ separating the plasma from the vacuum -- it is a FFS indeed. 
\par

In the vacuum,
\begin{eqnarray}
	E_{z} = \frac{4\pi\mu z}{3}\left(\frac{r_{0}^{3}}{r^{3}}-1\right) 
	\label{}
\end{eqnarray}
and
\begin{eqnarray}
	\frac{\partial E_{z}}{\partial z} = \frac{4\pi\mu}{3}
	\left(\frac{r_{0}^{3}}{r^{3}}-1-\frac{3r_{0}^{3}z^{2}}{r^{5}}\right) \,,
	\label{}
\end{eqnarray}
whence
\begin{eqnarray}
	\frac{\partial E_{z}}{\partial z}(r_{0},\theta) = -4\pi\mu\cos^{2}\theta
	= -4\pi\mu\cos^{2}\alpha \,:
	\label{}
\end{eqnarray}
we recover here in a particular situation 
our general relation (\ref{dEpar/dl}) with $\gamma=\gamma_{n}=1$.

\subsection{Nonexistence of voids in some particular $\B$}

Let $D_{v}$ be a void in contact with a force-free plasma region $D_{e}$ all along its boundary ${\cal S}$ -- a FFS. $D_{v}$ may be either of finite extent, or unbounded, in which case we also require that the electric field vanishes at infinity.
\par

We first consider the case where $\B$ is uniform ($\B=B\zu$). Then 
\begin{align}
	&\Na^{2}E_{z} = 0 &\mbox{in } D_{v} \,,
	\label{} \\
	&E_{z}=0 &\mbox{on } {\cal S} \,,
	\label{} \\
	&\lim_{r\to\infty}E_{z}=0 &
	\mbox{(if $D_{v}$ unbounded)} \,,
\end{align}
from which we can  conclude at once by a standard uniqueness theorem for Laplace equation that $E_{z}=-\dep{z}\psi=0$ in $D_{v}$. Therefore,
$\psi=f(x,y)$, with $f$ an harmonic function defined on the projection of $D_{v}$ onto the plane $(x,y)$. Because the potential is continuous on ${\cal S}$ and is constant along the lines in $D_{e}$, we must have $\psi(x,y,z)=f(x,y)$
and then $\mu=-\Na^{2}f/4\pi=0$ in that part of $D_{e}$ magnetically connected to $D_{v}$. This is in clear contradiction with the definition of
$D_{e}$, and we can conclude that the gap $D_{v}$ cannot exist. (Note that there is no disagreement between this result and the example of the previous subsection, as 
in the latter $E_{z}$ is not bounded when $r\to\infty$).
\par

Next we consider the case where $\B$ is radial  
($\B$ is the monopolar field $\B=B_{0}(r_{0}/r)^{2}\ru$, or a split monopolar field with zero flux, but with a current sheet) and $D_{v}$ is located in the region $r>r_{0}$, say. We note that the function $\chi=\r\cdot\Na\psi$ satisfies
\begin{align}
	&\Na^{2}\chi = 0 &\mbox{in } D_{v} \,,
	\label{} \\
	&\chi=0 &\mbox{on } {\cal S} \,,
	\label{} \\
	&\lim_{r\to\infty}\chi=0 &
	\mbox{(if $D_{v}$ unbounded)} \,,
\end{align}
which implies that $\chi=r\dep{r}\psi=0$ in $D_{v}$.
Then $\psi=g(\theta,\varphi)$ in $D_{v}$, with $g$ harmonic, and we can conclude by an argument similar to the one above that this relation also holds true in a part of $D_{e}$ -- a contradiction.
Vacuum gaps do not exist in the radial field either.

\end{document}